\documentclass[useAMS,usegraphicx,usenatbib]{mn2e} 
\usepackage{amsmath,graphicx}
\usepackage{subfigure}
\usepackage{ulem}
\usepackage{color}
\usepackage{graphicx}
\usepackage{times} 
\usepackage{amssymb}
\usepackage{amsmath}
\usepackage{lscape}
\usepackage{url}
\newif\ifAMStwofonts
\AMStwofontstrue

\def\rms{${\it r}_{\rm ms}$}
\def\rg{${\it r}_{\rm g}$}
\def\rin{${\it r}_{\rm in}$}
\def\laor{\rm{\sc LAOR}}
\def\phabs{\rm{\sc PHABS}}
\def\diskbb{\rm{\sc DISKBB}}
\def\refhiden{\rm{\sc REFHIDEN}}
\def\reflionx{\rm{\sc REFLIONX}}
\def\reflion{\rm{\sc REFLION}}

\def\kdblur{\rm{\sc KDBLUR}}

\def\nh{${\it N}_{\rm H}$}
\def\ka{$K\alpha$}

\def\epicpn{{\it EPIC}{\rm-pn}}
\def\epicmos1{{\it EPIC}{\rm-MOS1}}
\def\epicmos2{{\it EPIC}{\rm-MOS2}}
\def\epicmos{{\it EPIC}{\rm-MOS}}

\def\xmm{{\it XMM-Newton}}
\def\swift{{\it SWIFT}}
\def\asca{{\it ASCA}}
\def\integral{{\it INTEGRAL}}

\def\rxte{{\it RXTE}}

\def\xspec{\hbox{\sc XSPEC}}
\def\xspecv{{\sc XSPEC}{\rm\thinspace v\thinspace 12.4.0}}

\def\s{\hbox{$\rm\thinspace s$}}
\def\ks{\hbox{$\rm\thinspace ks$}}

\def\deg{$^{\circ}$}  

\def\cm{\hbox{$\rm\thinspace cm$}}

\def\kpc{\hbox{$\rm\thinspace kpc$}}

\def\ev{\hbox{$\rm\thinspace eV$}}
\def\kev{\hbox{$\rm\thinspace keV$}}

\def\msun{\hbox{$\rm\thinspace M_{\odot}$}}

\def\gx{\hbox{\rm GX 339-4}}
\def\jb{\hbox{\rm J1655-40}}
\def\ja{\hbox{\rm J1753.5-0127}}

\begin{document}

\title[Spin from disk reflection signatures] {Determining the spin of two stellar-mass black holes from disk
  reflection signatures } \author[R. C. Reis et al.]  {\parbox[]{6.in}
  {R.~C.~Reis $^{1}$\thanks{E-mail: rcr36@ast.cam.ac.uk}, A.~C.
    Fabian$^{1}$, R.~R. Ross$^{2}$ and J.~M. Miller$^3$\\ } \\
  \footnotesize
  $^{1}$Institute of Astronomy, Madingley Road, Cambridge, CB3 0HA\\
  $^{2}$Physics Department, College of the Holy Cross, Worcester, MA 01610, USA\\
  $^{3}$Department of Astronomy, University of Michigan, 500 Church
  Street, Ann Arbor, MI 48109, USA}

\maketitle

\begin{abstract}
  We present measurements of the dimensionless spin parameters and
  inner-disk inclination of two stellar mass black holes. The spin
  parameter of SWIFT \ja\ and GRO \jb\ are estimated by modelling the
  strong reflection signatures present in their \xmm\
  observations. Using a newly developed, self-consistent reflection
  model which includes the blackbody radiation of the disk as well as
  the effect of Comptonisation, blurred with a relativistic line
  function, we infer the spin parameter of SWIFT \ja\ to be
  $0.76^{+0.11}_{-0.15}$. The inclination of this system is estimated
  at $55^{+2}_{-7}$ degrees. For GRO{\thinspace \jb} we find that the
  disk is significantly misaligned to the orbital plane, with an
  innermost inclination of $30^{+5}_{-10}$ degrees. Allowing the
  inclination to be a free parameter we find a lower limit for the
  spin of $0.90$, this value increases to that of a maximal rotating
  black hole when the inclination is set to that of the orbital plane
  of \jb.  Our technique is independent of the black hole mass and
  distance, uncertainties in which are among the main contributors to
  the spin uncertainty in previous works.

\end{abstract}


\section{Introduction}

Black holes can be described by two observable parameters, mass and
spin. To date there are over twenty stellar mass black holes with
dynamically constrained mass (for a review see McClintock \& Remillard
2006); however for just a handfull of these systems do we have any
measurements of their spin.

Shafee et al. (2006) and McClintock et al. (2006) have reported values
of the dimensionless spin parameter, {\it a} for 4U 1543-47 of
0.7-0.85, GRO J1655-40 of 0.65-0.75, and GRS 1915+105 of 0.98-1. More
recently Liu et al. (2008) reported a value of $0.77\pm{\thinspace
  0.05}$ for M33{\thinspace X-7}. Their approach relies on modelling
the thermal continuum seen in the thermal X-ray spectra and requires
sources to be selected in the high-soft state (for a review on
spectral states see McClintock \& Remillard 2006. Any powerlaw
emission is then minimal and the spectrum resembles a quasi-blackbody
continuum. Furthermore, precise measurements of the mass and distance
to the black hole, as well as the inclination of the system, are
essential.

We have recently reported precise measurements of the spin parameter
in \gx\ (Reis et al. 2008; Miller et al. 2008a) using the reflection
signatures in the spectrum. These reflection features arise due to
reprocessing of hard X-ray by the cooler accretion disk (Ross \&
Fabian 1993), and consist of fluorescent and recombination emission
lines as well as absorption features. In the inner regions of an
accretion disk the various reflection features become highly distorted
due to relativistic effects and Doppler shifts. The shape of the
prominent Fe-\ka\ fluorescent line, and more importantly the extent of
its red wing, can give a direct indication of the radius of the
reflecting material from the black hole (Fabian et al. 1989, 2000;
Laor 1991). The stable circular orbit around a black hole extends down
to the radius of marginal stability, \rms, which depends on the spin
parameter (e.g Bardeen et al. 1972). A major advantage of using
reflection features to obtain the spin of the black hole is that these
features are completely independent of black hole mass and distance
and can thus be used for any system where both or either of these
parameters are unknown (for a review see Miller 2007). For \gx\ we
found that for both the very-high and the low-hard state the accretion
disk extends to the innermost stable circular orbit, \rms\ at a radius
of $\approx 2.05$\rg, where \rg$ = GM / c^{2}$ (Reis et al. 2008;
Miller et al. 2008a). This implies a spin of $\approx 0.935$ for \gx.

In this paper we use the method adopted by Reis et al. (2008) which
uses a specially developed spectral grid, \refhiden\ (Ross \& Fabian
2007), to obtain the spin parameter of {\rm SWIFT} \ja\ and {\rm
  GRO}{\thinspace\jb}. \ja\ was first detected in hard X-rays by
the Burst Alert Telescope (BAT) on the \swift\
satellite on 2005 May 30 (Palmer et al. 2005). Using \rxte\ and \xmm\
data, Miller, Homan \& Miniutti (2006, hereafter M06) showed the
presence of a cool ($kT \approx 0.2 \kev$) accretion disk extending
close to the \rms\ in the low-hard state (LHS) of the system. The
presence of this cool accretion disk was later confirmed by Ramadevi
\& Seetha (2007) and more recently by Soleri et al. (2008) using
multiwavelength observations of the source.

GRO{\thinspace}\jb\ was discovered by the Burst and Transient Source
Experiment (BATSE) on-board of the Compton Gamma Ray Observatory
(CGRO) on 1994 July 27 (Zhang et al. 1994). The mass of the compact
object has been dynamically constrained to $>6.0\msun$ (Orosz \&
Bailyn 1997). \asca\ observation of \jb\ from 1994 through 1996
provided the first clear detection of absorption-line features in the
source (Ueda et al. 1998; see also Miller et al. 2008b). Using
archival \asca\ data, Miller et al. (2004) showed evidence of highly
skewed, relativistic lines, and suggested an inner-radius of
reflection of $\approx 1.4$\rg\ which would indicate a rapidly
spinning black hole. This was later confirmed by Diaz-Trigo et
al. (2007, hereafter DT07) using simultaneous \xmm\ and \integral\
observations of \jb\ during the 2005 outburst. The high spin suggested
by these authors is in contrast with the spin parameter reported by
Shafee et al. (2006) of 0.65--0.75 using the thermal thermal continuum
method. In the following section we detail the observation, analyses
procedure and results.

\section{Observation and Data reduction}

\swift\ \ja\ was observed in its low-hard state by \xmm\ for 42\ks,
starting on 2006 March 24 16:00:31 UT and simultaneously by \rxte\ for
2.3\ks\ starting at 17:26:06 UT (M06). The \epicpn\ camera (Struder et
al. 2001) was operated in ``timing'' mode with a ``medium'' optical
blocking filter. For GRO \jb, observations were made by \xmm\ for
23.9\ks\ on 2005 March 18 15:47:13 (hereafter Obs 1) and again on 2005
March 27 08:43:59 (hereafter Obs 2) for 22.3\ks\ (DT07). The source
was found to be in the high-soft state and was observed with the
\epicpn\ camera in ``burst'' mode with a ``thin'' optical blocking
filter. Starting with the unscreened level 1 data files for all
aforementioned observation we extracted spectral data using the latest
{\it XMM-Newton Science analyses System v} 7.1.0 (SAS). For \ja\
events were extracted in a stripe in RAWX (20-56) and the full RAWY
range. RAWX 30--43 and RAWY 5--160 was used for \jb. For both sources,
bad pixels and events too close to chip edges were ignored by
requiring ``FLAG = 0'' and ``PATTERN$\le4$''. The energy channels were
grouped by a factor of five to create a spectrum. Background spectra
were extracted for both sources from an adjacent region of similar
RAWY and RAWX. In the case of \jb\ the background is negligible due to
the high source flux and it was not used for the analyses that
follows. In both cases, individual response files were created using
the SAS tools {\sc rmfgen} and {\sc arfgen}. The \epicpn\ resulted in
a total good-exposure time of 40.1\ks\ for \ja. Due to the low
duty-cycle of the ``burst'' mode (3\%) the total good exposure time
for \jb\ is $\approx 0.7$ and 0.6\ks\ for Obs 1 and 2
respectively. \rxte\ data for \ja\ were reduced in the standard way
using the {\it HEASOFT v 6.0} software package. We used the ``Standard
2 mode'' data from PCU-2 only. Standard screening resulted in a net
Proportional Counter Array (PCA) and High-Energy X-Ray Timing
Experiment exposures of 2.3 and 0.8\ks\ respectively. To account for
residual uncertainties in the calibration of PCU-2, we added 0.6 per
cent systematic error to all energy channels. The {\it HEXTE-B}
cluster was operated in the ``standard archive mode''. The {\it FTOOL}
{\sc grppha} was used to give at least 20 counts per spectral bin.

We restrict our spectral analyses of the \xmm\ \epicpn\ data for \ja\
to the 0.7--10.0\kev\ band. For \jb\ we use 0.7--9.0\kev\ due to the
uncertain calibration above 9.0\kev\ for the \epicpn\ burst mode. The
PCU-2 and HEXTE spectrum are restricted to 5.0--25.0 and
20.0--100.0\kev\ band respectively. A Gaussian line at 2.2--2.3\kev\
is introduced when fitting the \epicpn\ spectrum due to the presence
of an instrumental feature in this energy range that resembles an emission
line. This feature is likely to be caused by Au M-shell edges and Si
features in the detector. All parameters in fits involving different
instruments were tied and a normalisation constant was
introduced. \xspecv\ (Arnaud 1996) was used to analyse all
spectra. The quoted errors on the derived model parameters correspond
to a 90 per cent confidence level for one parameter of interest
($\Delta\chi^{2}=2.71$ criterion).

\section{analysis and results}

Fig. 1 shows the data/model ratio for \ja\ fitted with a simple
absorbed powerlaw (\phabs\ model in \xspec\ with \nh\ fixed at
$1.72\times10^{21} \cm^{-2}$) in the energy range 2.5--5.0 and
8.0--10.0\kev\ and then extended to the full energy range.  As first
noted by Miller, Homan \& Miniutti (2006) and more recently confirmed
by Soleri et al. (2008), \ja\ shows clear indications of the presence
of a cool accretion disk with a temperature of $\approx0.2$\kev\ in
its low-hard state. The presence of an excess at $\sim 6.9\kev$ is
indicative of Fe-\ka\ emission. Similarly for \jb, we fit the 0.7-4.0
and 7.0-9.0\kev\ energy range with a power law and a further
multi-color disk blackbody (MCD; Mitsuda et al. 1984) modified by
absorption in the interstellar medium. The value of \nh\ was kept
constant for the two observations. The data/model ratio for \jb\
extended to the full energy range is shown in Fig. 2. The presence of
a broad Fe-\ka\ line emission extending to just over 3.0\kev\ as well
as absorption features are clearly seen. Fitting the low/hard state of
\ja\ with a simple absorbed powerlaw results in an unacceptable fit
(Table 1) with various residuals in both the soft and the Fe-\ka\ energy
range.

The presence of a soft disk excess and broad Fe-\ka\ emission line is
usually modelled phenomenologically with a combination of a
multi-color disk component such as \diskbb\ (Mitsuda et al. 1984) and
a relativistic line such as the \laor\ line profile (Laor 1991). This
combination of components can give robust spin measurements when the
presence of the Fe-\ka\ emission line is significantly above the
continuum. However it should be noted that it is only an approximation
since the identification of \rin, as determined from \laor\ assumes a
hard wired spin parameter of $a=0.998$. Furthermore, the \laor\ line
profile is a phenomenological model for a relativistic {\it emission
  line}. The effects that extreme gravity have on the reflection
signatures are not limited to the Fe-\ka\ line profile and thus a more
thorough method to constrain the spin would involve modelling all of
the reflection signatures present in the spectra. In what follows, we
use the reflection model developed by Ross \& Fabian (2007, \refhiden)
to model all the reflection signatures as well as the disk
blackbody emission in a self-consistent manner. This approach was
detailed in Reis et al. (2008) for the galactic black hole \gx.

\begin{figure}

\rotatebox{270}{
\resizebox{!}{8.cm} 
{\includegraphics{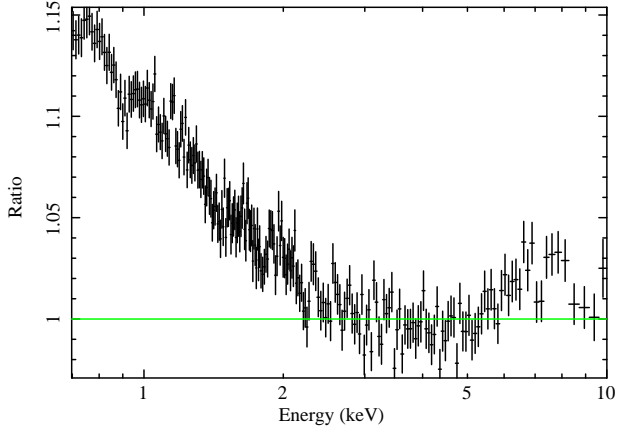}}
}
\caption{Data/model ratio for \ja\ obtained by fitting the energy
  range 2.5--5.0 and 8.0--10.0\kev\ with an absorbed powerlaw. It is
  clear that a semi-blackbody component as well as an iron reflection
  signature is present in the low-hard state of \ja. The data have
  been rebinned for plotting purposes only.}

\end{figure} 
\begin{figure}

\rotatebox{270}{
\resizebox{!}{8cm} 
{\includegraphics{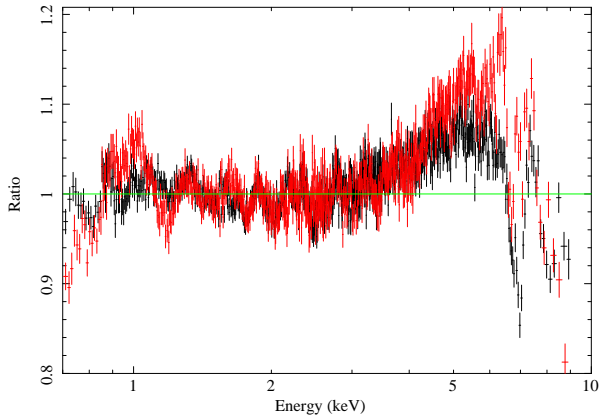}}
}
\caption{Data/model ratio for \jb\ using a simple absorbed power law
  and disk blackbody fitted in the energy range 0.7--4.0 and 7.0--9.0
  \kev. Obs 1 and 2 are shown in black and red respectively. The
  presence of a broad Fe-\ka\ line and complex absorption is clearly
  seen. The data have been rebinned for plotting purposes only.}

\end{figure}

The reflection features produced by the illuminated surface layer of
an accretion disk are largely dependent on the ionisation state of the
disk, and thus an important quantity is the ionisation parameter
$\xi=4\pi F_h/H_{den}$, where $F_h$ is the hard X-ray flux
illuminating a disk with a hydrogen density $H_{den}$ (Matt, Fabian \&
Ross 1993). The \refhiden\ reflection model incorporates the
importance of the thermal emission from the disk mid-plane in
determining the ionisation state and thus reflection features. Whereas
previous reflection models such as the Constant Density Ionised Disk
(CDID; Ballantyne, Ross \& Fabian 2001) and \reflionx\ (Ross \& Fabian
2005) vary over the ionisation parameter $\xi$ and does not account
for thermal emission, in \refhiden\ both the number density of
hydrogen, ${\it H}_{\rm den}$, and the ratio of the the total flux
illuminating the disk to the total blackbody flux emitted by the disk
($R_{illum/BB}$) are implicit parameters. The value of {\it kT} for
the blackbody entering the surface layer from below and the power-law
photon index are further parameters in the model. The disk reflection
spectra is convolved with relativistic blurring kernel \kdblur, which
is derived from the code by Laor (1991). The parameters for the
blurring kernel are the emissivity index $q_{in}$, where the
emissivity ($\epsilon_{r}$) is described by a powerlaw-profile such
that $\epsilon_{r}=r^{-q_{in}}$, disk inclination $i$, the inner disk
radius \rin, and the outer disk radius which was fixed at 400\rg. The
power law index of \refhiden\ is the same as that of the hard
component.

\begin{table}
\begin{center}
  \caption{Results of fits to \xmm\ \epicpn\ data for \ja\ with both a
    simple absorbed powerlaw and the self-consistent reflection model
    \refhiden.  }

\begin{tabular}{lcccccccccc}                
\hline
\hline
Parameter & Simple & \refhiden     \\
\hline
\nh\ $(10^{21} \cm^{-2})$ & $1.54\pm0.13$ &  $1.52^{+0.02}_{-0.03}$ \\
$\Gamma$ & $1.631\pm0.003$ &  $1.54^{+0.01}_{-0.04}$ \\
$N_{\rm PL}$ &  $0.0591\pm0.0002$ & $0.050^{+0.003}_{-0.004}$  \\
{\it kT} (\kev)& & $0.193^{+0.003}_{-0.002}$  \\
$H_{\rm den}(\times10^{20}{\thinspace {\rm H}\cm^{-3})} $& & $1.40^{+0.13}_{-0.27}$ \\
$R_{Illum/BB}$ && $9.6^{+0.4}_{-0.6}$ \\
$N_{\refhiden}$  && $0.0013^{+0.0004}_{-0.0006}$  \\
$q_{\rm in}$ && $4^{+6}_{-1}$ \\
\rin (\rg) && $3.1^{+0.7}_{-0.6}$  \\
{\it i} (\rm deg) & &$55^{+2}_{-7}$  \\
$\chi^{2}/\nu$ & $2191 .9/1859$ & $1839.5/1852$  \\
\hline
\hline
\end{tabular}
\end{center}

\small Notes.- The self-consistent model is described in \xspec\ as PHABS$\times$KDBLUR$\times$(PL+REFHIDEN). The normalisation of each component is referred to as {\it N}. All errors refer to the 90\% confidence range for a single parameter.

\end{table}

\subsection{Self-consistent reflection and disk emission }

$\bullet$ {\bf SWIFT J1753.5-0127}: The model provides an excellent
fit to the data for \ja\ with $\chi^{2}/\nu =1839.5/1852$. Table 1
details the parameter values with all errors corresponding to the 90
per cent confidence range. The data/model spectrum is shown in Fig. 3a
with the fit extended to 100.0\kev\ using \rxte\ data shown in the
inset. A normalisation constant was added to account for flux mismatch
between the instruments. The best fit model prior to gravitational
blurring is shown in Fig. 3b. The value of the inner radius obtained
from the gravitational blurring of the reflection features is
constrained to be \rin$ = 3.1^{+0.7}_{-0.6}$\rg. The strong constraint
on the value of \rin\ can be better appreciated in Fig. 4a, where the
90 per cent confidence level is shown in the $\chi^{2}$ plot obtained
with the ``steppar'' command in \xspec. Fig. 4b shows a similar
constraint obtained for the inner accretion disk inclination of \ja.

\begin{figure*}
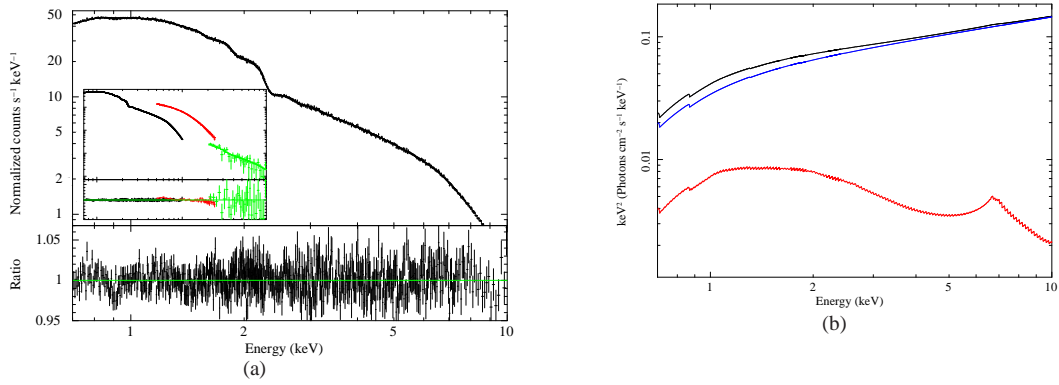

\centering
\subfigure[] 
{
 \rotatebox{270}{
\resizebox{!}{7.cm} 
{\includegraphics{figure_finalspec.ps}  
}}}
\hspace{1cm}
\subfigure[] 
{  
  \rotatebox{270}{
\resizebox{!}{6.cm} 
{\includegraphics{model.ps}} 
}}

\caption{{\rm (a)}: Data/model ratio for \ja\ in its low-hard
  state. The model assumes a powerlaw emissivity profile and
  constitutes of a powerlaw and the disk reflection model
  \refhiden. The inset shows the model extended to 100.0\kev\ using
  \rxte\ PCA (red) and HEXTE (green). Data have been rebinned for
  plotting purposes only. {\rm (b)}: Best-fit model prior to
  gravitational blurring showing the reflection features. The
  total model, powerlaw and reflection components are shown in black,
  blue and red respectively.}

\end{figure*}

\begin{figure*}
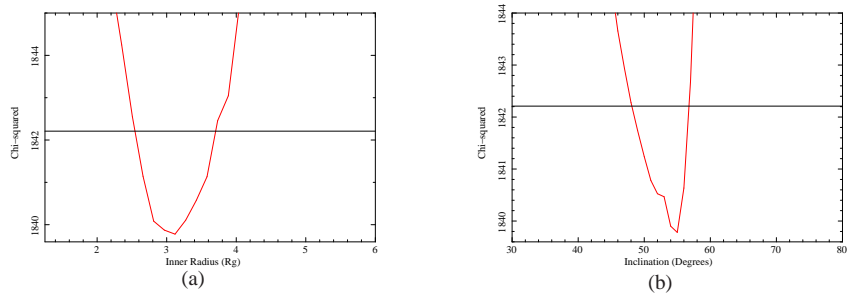

\centering
\subfigure[] 
{
 \rotatebox{270}{
\resizebox{!}{5.cm} 
{\includegraphics{steppar_rin_zoomedout.ps}  
}}}
\hspace{1cm}
\subfigure[] 
{  
  \rotatebox{270}{
\resizebox{!}{5.cm} 
{\includegraphics{steppar_inclinationzoomedout.ps}} 
}}
\caption{{\rm (a)}: $\chi^{2}$ vs \rin\ plot for \ja. A value of
  \rin$=3.1^{+0.7}_{-0.6}$\rg\ is found at the 90 per cent confidence
  level for one parameter of interest ($\Delta\chi^{2} = 2.71$
  criterion) shown by the solid horizontal line. {\rm (b)}: Similar plot
  for the disk inclination. A value for the disk inclination of
  $55^{+2}_{-7}$ is found at the 90 per cent confidence level for one
  parameter of interest. }

\end{figure*}

\begin{figure*}
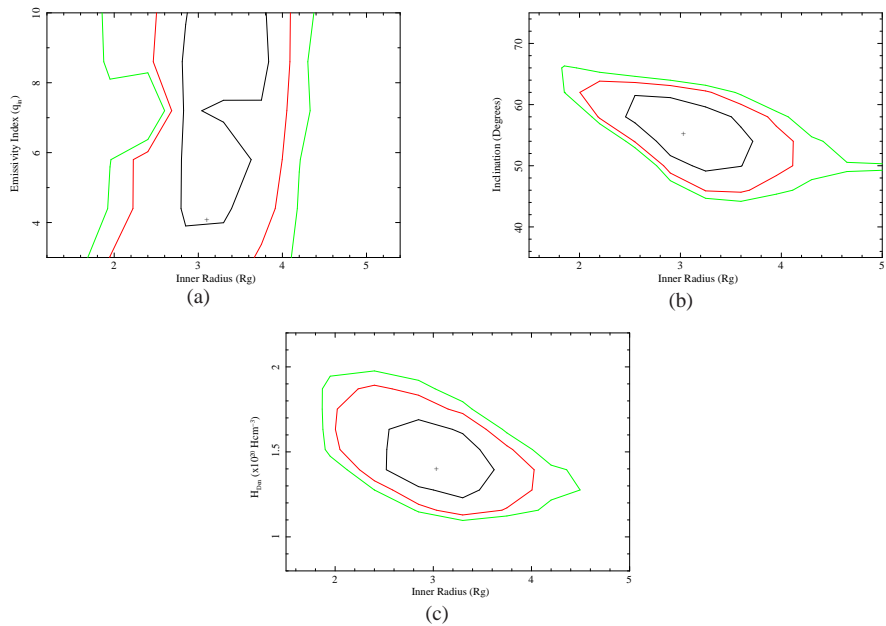

\centering
\subfigure[] 
{
 \rotatebox{270}{
\resizebox{!}{5.2cm} 
{\includegraphics{contour_q_rin.ps}  
}}}
\hspace{1cm}
\subfigure[] 
{
    
  \rotatebox{270}{
\resizebox{!}{5.2cm} 
{\includegraphics{contour_inc_rin.ps}} 
}}
\hspace{1cm}
\subfigure[] 
{
    
  \rotatebox{270}{
\resizebox{!}{5.2cm} 
{\includegraphics{contour_hden_rin.ps}} 
}}
\label{fig:sub} 
\caption{{\rm (a)}: Emissivity versus inner radius contour plot for \ja. The
  68, 90 and 95 per cent confidence range for two parameters of
  interest are shown in black, red and green respectively. {\rm (b)}:
  Similar plot for the inclination versus inner radius and {\rm (c)}:
  Hydrogen density versus inner radius.  It can be seen that for the
  full range of the emissivity, inclination and hydrogen density, the
  inner radius is constrained between approximately 2--4\rg\ at the 90
  per cent confidence level for two parameters. }

\end{figure*}

As can be seen from Table 1, the value of the emissivity index,
$q_{in}$ has been poorly constrained. A value of 3 is expected for a
standard accretion disk (Reynolds \& Nowak 2003) with steeper values
usually being interpreted as emission from a small, compact, centrally
located X-ray source, such as expected from the base of a jet
(Miniutti \& Fabian 2004). In order to investigate any degeneracy
between the value of the inner radius and the unconstrained emissivity
we explored their parameter space using the ``contour'' command in
\xspec. All parameters except for $\Gamma$ and $R_{Illum/BB}$ were
free to vary. Fig. 5a shows the 68, 90 and 95 per cent contour for two
parameters of interest. Although the emissivity index is poorly
constrained, it can be seem from Fig. 5a that the value of the inner
radius is not strongly affected by this uncertainty. In what follows
we will thus freeze the value of the emissivity at the best fit value
shown in Table 1. Fig. 5b and 5c shows similar contour plots for both
inclination and $H_{den}$ versus inner radius respectively. For a
large range of inclination and $H_{den}$ the value of \rin\ remains
approximately between 2--4\rg\ with a best fit value of approximately
$3$\rg, in accordance to that shown in Fig. 4a. Assuming that \rin$ =
3.1^{+0.7}_{-0.6}$\rg\ (Fig. 4a) is the same as the radius of marginal
stability \rms, we obtain a dimensionless spin parameter of $
0.76^{+0.11}_{-0.15}$ for \ja\ in its low-hard state (Fig. 6).

\begin{figure}
\centering
{
{\includegraphics[height=5.5cm, width=7.5cm]{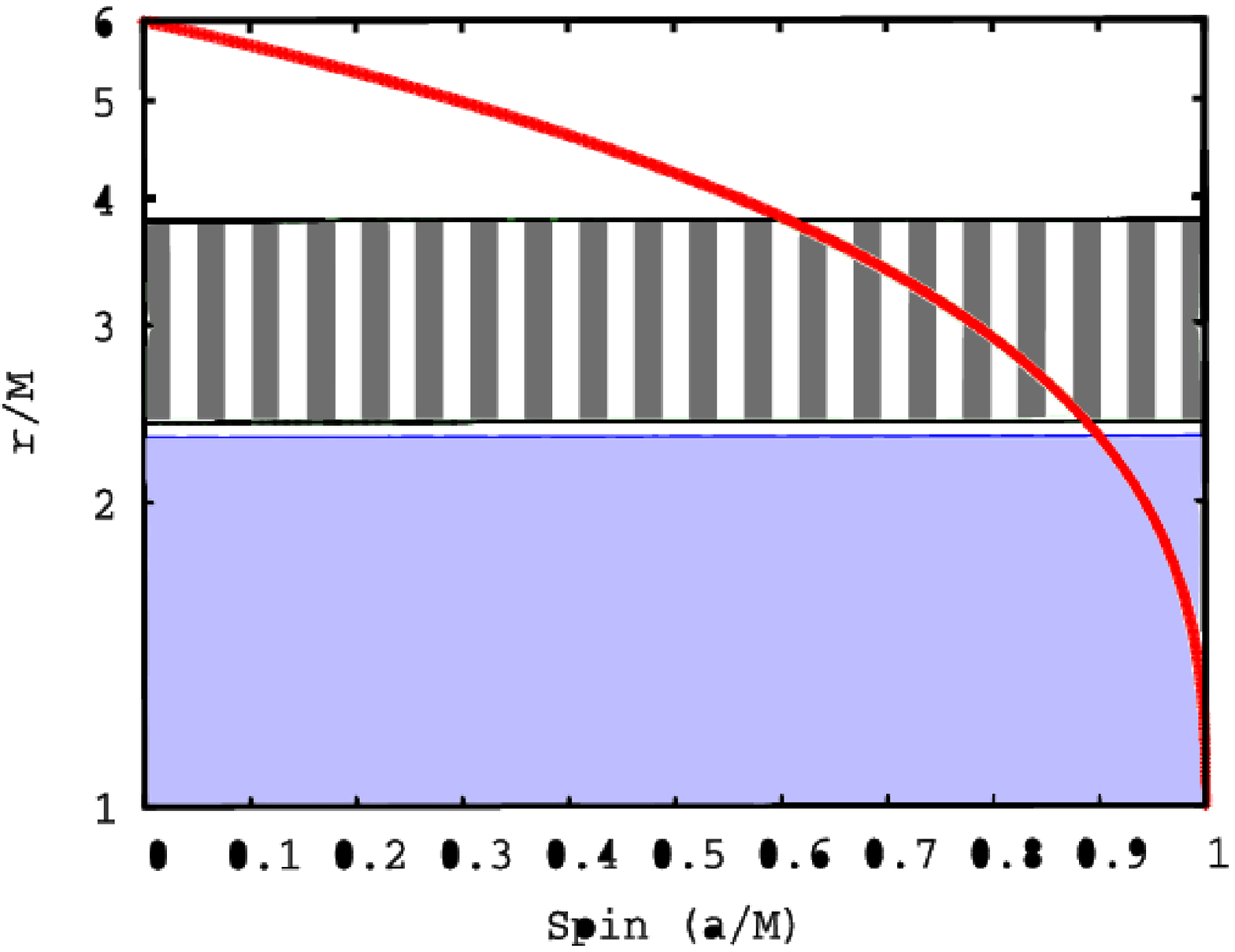}}
}
\vspace{0.cm}
\caption {Plot of the innermost stable circular orbit vs dimensionless
  spin parameter (solid curve). The constraints imposed by the
  innermost radius obtained for \ja\ are shown by the intersection of
  the solid line with the vertical region. The solid region shows
  similar constrains for the lower limit of the spin parameter in
  \jb. Based on this analysis, a dimensionless spin parameters of
  $0.76^{+0.11}_{-0.15}$ is found for \ja\ and a lower limit of
  $\approx0.90$ is found for \jb.}

\end{figure}

\begin{table*}
\begin{center}
  \caption{Results of fits to \xmm\ \epicpn\ data for \jb\ with self-consistent reflection model \refhiden.  }

\begin{tabular}{lccccccccc}                
  \hline
  \hline
  Parameter & Model 1   &      & Model 2  &    & Model 3   &    & Model 4 &     \\
  & Obs 1 & Obs 2  & Obs 1& Obs 2& Obs 1& Obs 2 & Obs 1& Obs 2\\

  \nh & $0.76^{+0.01}_{-0.02}$&...&  $0.759\pm0.001$  &...&$0.765\pm0.001$&...& $0.742^{+0.004}_{-0.002}$&... \\
  $\Gamma$ & $2.59^{+0.03}_{-0.01}$&$2.88^{+0.03}_{-0.02}$ & $2.72\pm0.08$ & $2.91^{+0.04}_{-0.053}$&  $2.54^{+0.02}_{-0.01}$ & $2.99^{+0.01}_{-0.02}$&  $2.54^{+0.06}_{-0.02}$ & $2.69^{+0.03}_{-0.02}$      \\
  $N_{\rm PL}$ & $10.3^{+0.2}_{-0.4}$ & $10.4^{+0.3}_{-0.1}$&  $9.0^{+0.1}_{-0.2}$ & $11\pm3$& $10.4^{+0.1}_{-0.2}$ & $9.46^{+0.23}_{-0.02}$& $8.18^{+0.16}_{-0.12}$ & $9.35^{+0.14}_{-0.06}$ \\
  {\it kT} (\kev) & $0.846^{+0.002}_{-0.007}$ & $0.84\pm0.01$ & $0.778\pm0.009$&$0.848^{+0.003}_{-0.002}$& $0.776^{+0.002}_{-0.008}$&$0.780^{+0.009}_{-0.001}$& $0.804\pm0.004$&$0.816^{+0.007}_{-0.006}$ \\
  $H_{\rm den}$& $9.97^{+0.03}_{-0.56}$&$10^{+0.0}_{-0.2}$&  $9.0^{+0.3}_{-0.7}$&$8^{+1}_{-2}$& $7.4^{+0.3}_{-0.1}$&$10^{+0}_{-1}$& $10.0^{+0.0}_{-0.6}$&$9.0^{+0.7}_{-0.6}$ \\
  $R_{Illum/BB}$ & $0.93^{+0.02}_{-0.04}$&$0.20^{+0.03}_{-0.01}$ & $0.39^{+0.05}_{-0.10}$&$0.6\pm0.2$& $0.11^{+0.01}_{-0.03}$&$0.01^{+0.20}$ &$1.0^{+0.03}_{-0.02}$&$0.169^{+0.074}_{-0.007}$ \\
  $N_{ref}$  &$3.3^{+0.4}_{-0.2}$&$0.29^{+0.06}_{-0.09}$  &$1.4^{+0.1}_{-0.3}$&$0.7\pm 0.2$ & $0.46^{+0.31}_{-0.02}$&$0.02^{+0.10}$ &   $3.42^{+0.01}_{-0.04}$&$0.251^{+0.002}_{-0.007}$            \\
  $q_{\rm in}$ & $10.0_{-0.3}$&$4.5\pm0.1$    & $10.0^{+0.0}_{-0.3}$&$10.0^{+0.0}_{-0.5}$   & $10.0^{+0.0}_{-0.4}$&$7.75^{+0.06}_{-0.20}$    & $3.87\pm0.08$&$2.75^{+0.03}_{-0.02}$                        \\
  \rin (\rg) & $1.86^{+0.20}_{-0.02}$& ...  & $1.31\pm0.01$& ... & $1.38\pm0.01$& ...&$2.17^{+0.15}_{-0.17}$ &...\\
  {\it i}  & $50\pm1$&... &  70(f)&... &70(f)&...   &$30^{+5}_{-10}$&...  \\
  $E_{Gabs1}$ (\kev)& ...&...& ...&...&6.7(f)&6.7(f)&6.7(f)&6.7(f)\\
  $\sigma$(\kev) &...&...& ...&...&$0.08\pm0.02$&$0.003\pm0.001$& $0.09\pm0.03$&$0.004\pm0.001$      \\
  $\tau$ &...&...& ...&...&$0.030^{+0.003}_{-0.005}$&$0.20\pm0.06$ & $0.031^{+0.006}_{-0.005}$       &$0.22^{+0.08}_{-0.07}$  \\
  $E_{Edge1}$ (\kev)&...&...& ...&...&8.8(f)&8.8(f)&8.8(f)  &8.8(f)\\
  $E_{Gabs2}$ (\kev)&...&...& ...&...&6.97(f)&6.97(f)&6.97(f)&6.97(f)\\
  $\sigma$(\kev)&...&...& ...&...&$0.06\pm0.02$&$<0.8$& $0.056^{+0.019}_{-0.020}$& $<0.05$ \\
  $\tau$ & ...&...& ...&...&$0.041^{+0.004}_{-0.005}$&$0.014^{+0.016}_{-0.008}$ &$0.039^{+0.005}_{-0.002}$&$0.013^{+0.023}_{-0.005}$\\
  $E_{Edge2}$ (\kev)&...&...& ...&...&9.3(f)&9.3(f)&9.3(f)&9.3(f)   \\

  $\chi^{2}/\nu$&4967.3/3296&...&3230.0/2264&....&4425.3/3289&...& 4370.1/3288 \\

\hline
\hline
\end{tabular}
\end{center}

\small Notes.-Model 1 is described in \xspec\ as PHABS$\times$KDBLUR$\times$(PL+REFHIDEN). The value of  \nh, inclination and \rin\ were tied between the two observations. An instrumental line at $1.876$\kev\ was added to each model. The normalisation of each component is referred to as {\it N}. Frozen values are followed by (f). $H_{den}$ is given in units of $10^{21}{\thinspace {\rm H}\cm^{-3}}$. Model 2 is similar to the previous model and only fitted to the data below 6.6\kev. Model 3 and 4 includes Fe XXV and Fe XXVI absorptions ({\rm GABS} model in \xspec) at fixed energies of 6.7 and 6.97\kev\ respectively. Their respective absorption edges is modelled with the model {\rm EDGE} in \xspec\ with energies frozen at 8.8 and 9.3\kev\ respectively. The optical depth $\tau$ of the absorptions and edges are linked for consistency. 

\end{table*}

\vspace{1cm}

$\bullet$ {\bf GRO J1655-40}: Strong absorption features are clearly
present in the spectra of \jb\ (see Fig. 2), and thus a fit with
\refhiden\ should not immediately give an acceptable result. Fig. 7
(Top) shows the best-fit data/model ratio using \refhiden. The various
parameters for this model are described in Table 2 (Model 1). It can
be seen that, whereas in Fig. 2 there is evidence of both a broad line
and various absorption features, this time the presence of the broad
line has diminished (Fig. 7 Top). This fit constrains the inner radius
to \rin$=1.86^{+0.20}_{-0.02}$ \rg\ with $\chi^{2}/\nu
=4967.3/3296$. The best fit using Model 1 seems to imply a relatively
low inclination of 50\deg$\pm1$\deg, considerably less than the value
of the binary inclination of $70.2$\deg$\pm1.9$\deg\ (Greene, Bailyn
\& Orosz 2001). However, this value is still in agreement with that
presented by Diaz-Trigo et al. (2007, see their Table 1). The low
inclination seen here could be due to the various absorption features
masking themselves as the blue wing of an iron-\ka\ line
profile{\footnote {The blue wing of the iron-\ka\ line profile can be
    used to determine the inclination of the source (see e.g. Reynolds
    \& Nowak 2003; Fabian \& Miniutti 2005)}}.  Although various
features contribute to the determination of the inner radius of
emission and thus spin of the black hole, the most important of those
features for the purpose of this work is the extent of the red wing in
the Fe-\ka\ fluorescent line. In order to explore whether the
absorption features are affecting the value of the inner radius, we
froze the inclination at 70 degrees and modelled the spectra below
6.4\kev. The results of this fit is detailed in Model 2 (Table 2) and
shown in Figure 7 (Bottom). By restricting the inclination to the
known value of the binary inclination ($\approx70$ degrees) and
fitting the spectra below 6.4\kev\ the innermost radius approaches
that of a maximally rotating black hole with \rin$=1.31\pm 0.01$\rg\
and $\chi^{2}/\nu =3230.0/2264$. The majority of the contribution to
$\chi^{2}$ is now coming from residuals between 0.9--1.1\kev\ possibly
due to the photo-ionisation edge of O VIII at $\approx 0.9$\kev. In
order to extend this fit to the full range we modelled the various
absorption features in a phenomenological manner using two negative
Gaussian absorptions (GABS model in \xspec) fixed at energies of 6.7
and 6.97\kev\ corresponding to absorption of Helium-like Fe-$\rm XXV$
and Hydrogen-like Fe-$\rm XXVI$ respectively as well as absorption
edges fixed at 8.8 and 9.3\kev. The optical depths $\tau$ of the
absorptions and edges were linked for self-consistency. The various
parameters for this model covering the full energy range are detailed
in Table 2 (Model 3). The value of the inner radius remains low
($<1.4$\rg) similarly to that of Model 2. Allowing the inclination to
vary over its full range improved the statistics with
$\Delta\chi^{2}=-55.2$ for one less degree of freedom (Model 4) and
resulted in an inner radius of \rin$=2.17^{+0.15}_{-0.17}$\rg\
(Fig. 8a) and inclination of $30$\deg$^{+5}_{-10}$ (Fig. 8b). The best
fit spectra for \jb\ including the absorption lines and edges is shown
in Figure 9. Note that as we remove the restriction on the
inclination, the emissivity profile of the two observations become
less steep, with the indices constrained to $q_{in}=3.87\pm0.08$ and
$2.75^{+0.03}_{-0.02}$ for obs 1 and 2 respectively. Adding a possible
nickel emission line at $7.48$\kev\ with equivalent width of 30 and
17\ev\ for obs 1 and 2 respectively resulted in an improved fit with
$\Delta\chi^2=-72.2$ for 4 extra degrees of freedom. As was the case
for Model 2, most of the contribution to chi-squared are now coming
from residuals in the soft energy range. By restricting this fit to
above 1.5\kev\ we obtain $\chi^{2}/\nu =3496.0/2957$.

\begin{figure}
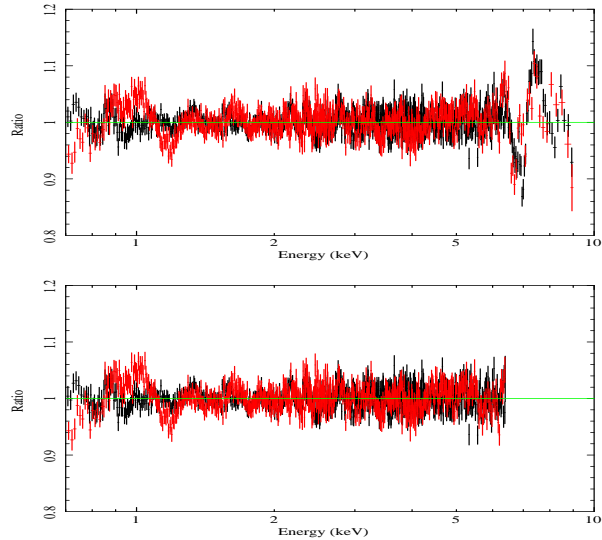


\rotatebox{270}{
{\includegraphics[width=35mm,height=80mm]{ratioplot_simple.ps}}
}
\rotatebox{270}{
{\includegraphics[width=35mm,height=80mm]{ratioplot_below64.ps}}
}

\caption {{\it Top:} Best-fit data/model ratio for \jb\ using Model 1
  (Table 2). The various absorption features are clearly seen. {\it
    Bottom:} Similar as above for Model 2 covering the energy range
  0.7--6.4\kev. The model clearly results in a good fit for the
  red-wing of the iron-\ka\ fluorescent line. Obs 1 and 2 are shown in
  black and red respectively. The data have been rebinned for plotting
  purposes only. }

\end{figure}

\begin{figure}
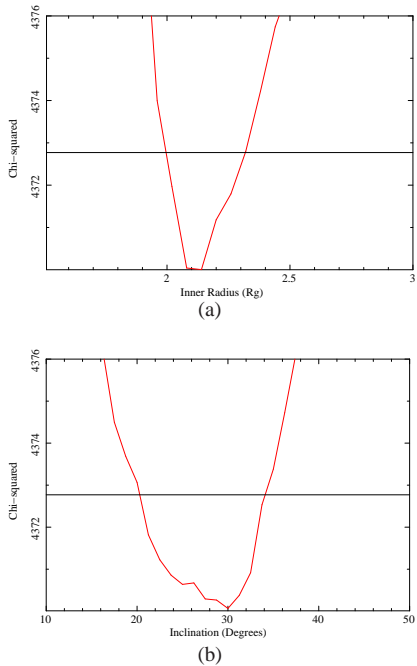

\centering
\subfigure[] 
{
 \rotatebox{270}{
\resizebox{!}{5.5cm} 
{\includegraphics{steppar_rin_j1655.ps}  
}}}
\hspace{1cm}
\subfigure[] 
{  
  \rotatebox{270}{
\resizebox{!}{5.5cm} 
{\includegraphics{steppar_inclination_j1655.ps}} 
}}

\caption{{\rm (a):} $\chi^{2}$ vs \rin\ plot for \jb. Using Model 4
  (Table 2) with the inclination allowed to vary over its full range
  we obtain a value of $2.17^{+0.15}_{-0.17}$\rg\ for the inner
  radius at the 90 per cent confidence level for one parameter of
  interest ($\Delta\chi^{2} = 2.71$ criterion) shown by the solid
  horizontal line.{\rm (b):} Similar plot for the inclination of \jb. }

\end{figure}

\begin{figure}

\rotatebox{270}{
\resizebox{!}{8cm} 
{\includegraphics{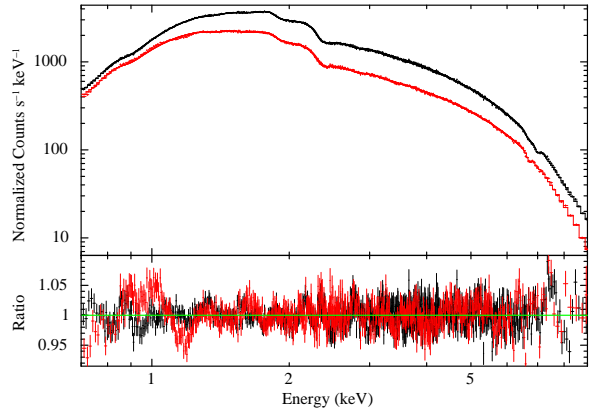}}
}
\caption {Best-fit spectra for \jb\ using Model 4. Obs 1 and 2 are
  shown in black and red respectively. The complex absorptions are
  phenomenologically modelled with two Gaussian absorption (see
  text). The data have been rebinned for plotting purposes only.}

\end{figure}

Similarly to \ja\ we investigate any degeneracy between the
inclination and inner radius in Model 4 by using the ``contour''
command in \xspec. All parameters except for $\Gamma$ and
$R_{Illum/BB}$ were free to vary. Fig. 10 shows the 68, 90 and 95 per
cent contour for the two parameters of interest. It can be seem from
Fig. 10 that for a large range of inclination the value of \rin\
remains below 2.35\rg\ with a best fit value of approximately
$2.15$\rg, in agreement with the result presented in Fig. 8a.  It is
clear that the unknown inclination of the system plays an important
role in the determination of the inner radius (and thus spin) of \jb,
with values as low as 1.3\rg\ (spin $>0.99$) obtained with an
inclination of 70 degrees (Model 2) rising to 2.32\rg\ (spin $>0.9$)
with the inclination allowed to vary (Model 4). However, it must be
noted that using the reflection model both with and without accounting
for the absorption features as well as both with and without
constraints on the inclination, gives results with an inner-radius
consistently below 2.32\rg\ (Table 2). Taking this value as an upper
limit to the innermost radius of emission implies a dimensionless spin
parameter greater than 0.9 for \jb\ (Fig. 6).

\begin{figure}

\rotatebox{270}{
\resizebox{!}{8cm} 
{\includegraphics{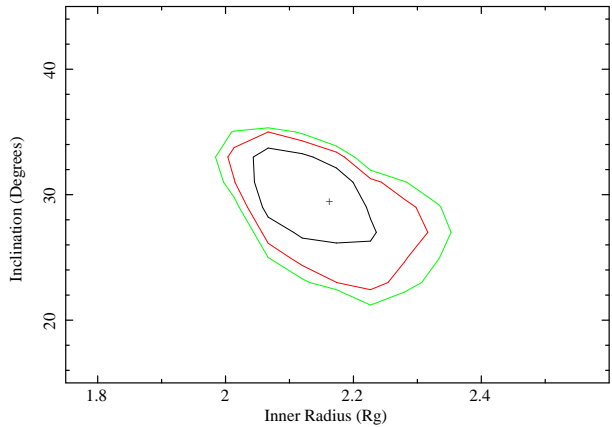}}
}
\caption {Inclination-inner radius contour plot for \jb. The 68, 90
  and 95 per cent confidence range for two parameters of interest are
  shown in black, red and green respectively. It can be seen that for
  the inner radius is constrained to be below approximately 2.5\rg for
  a large range of inclination at the 90 per cent confidence level for
  two parameters.}

\end{figure}

\section{Discussion}

The X-ray spectra of galactic black hole binaries can provide
important information on both the geometry of the system as well as
intrinsic physical parameters such as the spin of the central black
hole. Using the reflection features embedded in the spectra of both
\ja\ and \jb\ we found clear indications that the accretion disk
extends close to the radius of marginal stability in both cases. For
\ja\ we have shown that the innermost emitting region extends down to
\rin$= 3.1^{+0.7}_{-0.6}$\rg\ (Fig. 4a) with an innermost inclination
of $55^{+2}_{-7}$ degrees (Fig. 4b). Based on the normalisation of the
disk blackbody component, Miller, Homan \&\ Miniutti (2006) found an
inner radius of \rin$=2.0(3)(M/10\msun)(d/8.5\kpc)(cos
i)^{-1/2}$\rg, similar to ours for a large range of parameters. Our
result is also consistent with that found by Soleri et al. (2008) of
2.6--6.0\rg\ for a 10\msun\ black hole. In both these cases \rin\ was
estimated using the normalisation of the MCD component and thus
requires knowledge of the mass, distance and inclination of the
sources. Uncertainties in these values can significantly affect the
value of the inner radius and thus spin parameter.

Estimating the spin of a black hole based on the extent of the
emitting region relies on the assumption that the accretion disk
extends to the innermost stable circular orbit (ISCO) and that
emission within this radius is negligible. Reynolds \& Fabian (2008)
addressed the robustness of this assumption and found that reflection
within the ISCO becomes significantly less as one considers more
rapidly rotating black holes. The way the dimensionless spin parameter
depends on the position of the ISCO is shown in Fig. 6 (solid curve). Assuming that the accretion disk extend to the ISCO for \ja\ we
obtain a spin parameters of $0.76^{+0.11}_{-0.15}$. Note that this
estimate is independent of the unknown distance to the source.

The system in \jb\ is known to have an orbital-plane inclination of
approximately 70 degrees (Greene, Bailyn \& Orosz 2001). Constraining
the inclination to this value resulted in an inner radius of emission
consistent with that of a maximally rotating black hole (Model 3,
Table 2). Allowing the inclination of the blurring function to be a
free parameter in the spectral fit of \jb\ we obtain an improved fit
with an inner radius of $=2.17^{+0.15}_{-0.17}$\rg\ (Fig. 8a) and an
inner-disk inclination of $30^{+5}_{-10}$ degrees (Fig. 8b). This
value for the inner radius implies a {\it lower limit} for the spin
parameter of 0.9 (Fig. 6a). The possibility of a misalignment between
the innermost region of the accretion disk and the orbital plane
inclination in galactic microquasars have been suggested by various
authors (see e.g. Maccarone 2002) with clear examples seen in both
\jb\ (Martin, Tout \& Pringle 2008) and V4641 Sgr (Martin, Reis \&
Pringle 2008). It was shown in Martin, Tout \& Pringle (2008) that the
alignment time scale in microquasar such as \jb\ is usually a
significant fraction of the lifetime of the system. Thus if the black
hole in such a system were formed with misaligned angular momentum, as
expected from supernova-induced kicks, then it would be likely that
the system would remain misaligned for most of its lifetime. This
assumption is supported by the apparent misalignment found in the
present work.

The lower limit for the spin parameter of \jb\ found here ($>$0.9;
Fig. 6) is not consistent with that reported by Shafee et al. (2006)
of 0.65--0.75. As was mentioned above their method requires prior
knowledge of several other factors including the inner disk
inclination and distance to the source. Pszota \& Cui (2007) have
recently shown that neither disk continuum models {\it KERRBB} (Li et
al. 2005) nor {\it BHSPEC} (Davis \& Hubeny 2006) were able to
successfully model the ultrasoft spectra of \gx, nonetheless their
best fit parameters in conjunction with the best estimates for the
physical parameters of the source, suggested a moderate spin of
0.5--0.6 in comparison with our estimate of $\approx 0.935$ (Reis et
al. 2008; Miller et al. 2008) based on the various reflection features
in the spectra of \gx. Our suggestion that \jb\ contains a rapidly
rotating black hole is in agreement with results based on
Quasi-Periodic Oscillations where a value of $>0.91$ is usually
derived (Zhang et al. 1997; Cui et al. 1998; Wagoner et al. 2001;
Rezzolla et al. 2003). Furthermore, in their derivation of the spin
parameter of \jb, Shafee et al. (2006) assumed both a lack of
misalignment between the central disk region and the orbital plane and
more importantly they used a distance to the source of
$3.2\pm0.2$\kpc. Foellmi (2008) has recently shown that the distance
to \jb\ is most likely to be less than 2\kpc. When this distance is
used in place of 3.2\kpc\ the spin parameter of \jb\ derived by the
thermal continuum method becomes greater than 0.91 in agreement with
our results.

\section{Conclusions}

We have studied the \xmm\ spectra of both SWIFT \ja\ and GRO \jb. By
estimating the innermost radius of emission in these systems we
constrain their spin parameters by assuming that the accretion disk
extends down to the radius of marginal stability.  For \ja\ the spin
is found to be $0.76^{+0.11}_{-0.15}$ at 90 per cent confidence. The
innermost disk inclination in \ja\ is estimated at $55^{+2}_{-7}$
degrees. In the case of \jb\ we find that the best fit requires a disk
which is significantly misaligned to the orbital plane. An inclination
of $30^{+5}_{-10}$ degrees and a dimensionless spin parameter greater
than 0.9 is found at the 90 per cent confidence level. Our method
involves spectral modelling of both the intrinsic disk thermal
emission as well as the reprocessed hard radiation which manifest
itself as various reflection ``signatures''. These reflection
signatures are independent of both black hole mass and distance and
are thus a very useful tool for such measurements.

\section*{Acknowledgements}

RCR acknowledges STFC for financial support. ACF and RRR thanks
the Royal Society and the College of the Holy Cross respectively. 

\vspace{0.5cm}

\noindent {\bf Note added after submission:} In a recent arXiv
posting, Hiemstra et al. (2009) published an analysis made on the same
\xmm\ and \rxte\ data for \ja\ used here. They confirm the existence
of a broad iron-\ka\ line and show that various continuum models with
the addition of a \laor\ line can successfully fit the data. More
importantly, they claim to successfully fit the X-ray spectrum of \ja\
with a continuum model that does not require emission from a disk
extending down to the ISCO. We have investigated this solution and
find that in their case the broad line in a consequence of Compton
broadening of the emitted photons by the hot surface of the accretion
disk (Ross, Fabian \& Young 1999; Ross \& Fabian 2007). Their solution
requires a highly ionised disk ($\xi \sim 5000$ ${\rm erg}\cm\s^{-1})$
truncated at $\sim255$ \rg\ with an inclination angle consistent with
zero.  Using the same reflection model (\reflion; Ross \& Fabian 2005)
convolved with \kdblur\ and including a disk component, we find a
further solution where the disk extends down to the ISCO (6\rg) and
has a much lower ionisation parameter of $\xi \sim 500$ ${\rm
  erg}\cm\s^{-1}$. This second interpretation yields an inclination of
$\sim 65$ degrees, and an improvement of -7.7 in $\chi^2$ for 2 extra
degrees of freedom. The solution presented in our paper reflects that
of the lower ionisation reported above. We are concerned that in a
solution with such high ionisation parameter and a truncation radius
at $\sim 45$ times the ISCO would result in a disk which is
approximately $2\times 10^4$ times less dense that that of a similar
disk extending down to the ISCO. With an estimated mass of
approximately 10\msun\ (Cadolle Bel et al. 2007), a radius of 255\rg\
is the equivalent of a disk truncated at $\approx 3.8\times 10^8$\cm\
from the central black hole. Using the ionisation parameter and
unabsorbed flux presented in Hiemstra et al. (2009) we can estimate
the hydrogen number density to be approximately $2.3\times 10^{15}{\rm
  cm}^{-3}$. A modest surface layer of Thomson depth $\tau_T \sim 3$
will thus results in a disk with a half-thickness, $t$ greater than
$2\times 10^9 {\rm cm}$; at least 5 times larger than the truncation
radius. For a disk with inner radius $r>>$\rg\ we can write the
ionisation parameters as $\xi \approx 1.13\times10^5 \eta\ t/r^2 { \rm
  erg}\cm\s^{-1}$, where $\eta$ is the X-ray efficiency (Reynolds \&
Begelman 1997). Fig. 11 shows the ratio $t/r$ as a function of $r$ for
a system with $\eta=0.06$ and an ionisation parameter of 5000 and 500
${\rm erg}\cm\s^{-1}$. For a typical
thin accretion disk $t/r$ should be well below unity and as can be
seen in Fig. 11 there are no solutions in this range for a disk with
$\xi \sim 5000$ ${\rm erg}\cm\s^{-1}$. We conclude that the high
ionisation solution found by Hiemstra et al (2009) is physically
inconsistent; the disc must extend in to small radii.

\begin{figure}

\rotatebox{270}{
\resizebox{!}{8cm} 
{\includegraphics{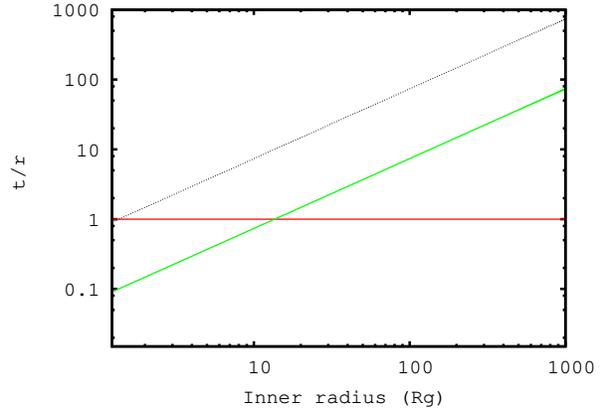}}
}
\caption { Ratio of half-thickness $t$ over inner radius $r$ as a
  function of $r$ for $\xi=5000$ (dotted line) and $\xi= 500$ ${\rm
    erg}\cm\s^{-1}$ (solid line). For a typical thin accretion disk
  this ratio should be well below 1 (horizontal line). }

\end{figure}


\begin{thebibliography}{}
     
\bibitem{} Arnaud, K.A., 1996, ASPC, 101, 17A
	
\bibitem{} Ballantyne, D. R., Ross, R. R., Fabian, A. C, 2001, MNRAS, 327, 10

\bibitem{} Bardeen, J.M., Press, W.H. \& Teukolsky, S.A., 1972, ApJ, 178,
  347

\bibitem{} Cadolle Bel, M., Rib{\'o}, M., Rodriguez, J., Chaty, S.,
Corbel, S., Goldwurm, A., Frontera, F., Farinelli, R., D'Avanzo, P.,
Tarana, A., Ubertini, P., Laurent, P., Goldoni, P., Mirabel, I.~F.,
2007, ApJ, 659, 549C



\bibitem{} Cui W., Zhang, S.~N., and Chen, W., 1998, ApJ, 493, L75

\bibitem{} Davis S.~W., and Hubeny, I., 2006, ApJS, 164, 530

\bibitem{} D'Angelo, C., Giannios, D., Dullemond, C., Spruit, H.,
  2008, A\&A, 488, 441

\bibitem{} D{\'{\i}}az Trigo, M., Parmar, A.~N., Miller, J., Kuulkers,
  E., and {Caballero-Garc{\'{\i}}a}, M.~D., 2007, AIPC, 924, 877
  (DT07)

 
\bibitem{} Fabian, A. C., Rees, M. J., Stella, L., White, N. E., 1989, MNRAS, 238, 729

\bibitem{} Fabian, A. C., Iwasawa, K., Reynolds, C. S., Young, A. J.,
  2000, PASP, 112, 1145
	
\bibitem{} Fabian, A. C., Miniutti, G., 2005, arXiv:astro-ph/0507409v1
\bibitem{} Foellmi, C., 2008, arXiv:0812.4232v1 [astro-ph]


\bibitem{} Greene, J., Bailyn, C. D., and Orosz, J. A., 2001, Apj,
  554, 1290

\bibitem{} Hiemstra, B., Soleri, P., Mendez, M., Belloni, T., Mostafa, R., Wijnands R., 2009, arXiv0901.2255H

\bibitem{} Laor, A., 1991, ApJ, 376, 90

\bibitem{} Li, L.-X., Zimmerman, E.~R., Narayan, R., and McClintock,
  J.~E., 2005, ApJS, 157, 335

\bibitem{} Liu, J., McClintock, J.~E., Narayan, R., Davis, S.~W. and
  Orosz, J.~A., 2008, ApJ, 679L, 37L

		
\bibitem{} Maccarone, T. J., 2002, MNRAS, 336, 1371

\bibitem{} Martin, R. G., Reis, R. C., Pringle, J. E., 2008, MNRAS, 391, 15	
\bibitem{} Martin, R. G., Tout, C. A., Pringle, J. E., 2008, MNRAS, 387, 188


\bibitem{} Matt, G., Fabian, A. C., Ross, R. R., 1993, MNRAS, 264, 839

\bibitem{} McClintock, J.E., \& Remillard, R.A. 2006 Black hole
  binaries (Compact stellar X-ray sources), 157-213

\bibitem{} McClintock, J.~E., Shafee, R., Narayan, R., Remillard, R.~A., Davis, S.~W., and Li, L.-X., 2006, ApJ, 652, 539


\bibitem{} Miller, J. M., Fabian, A. C., Nowak, M. A., Lewin,
  W. H. G., 2004, Proc. 10th Anual Marcel Grossmann Meeting, General
  Relativity. (astroph/ 0402101)

\bibitem{} Miller, J. M., 2007, ARA\&A, 45, 441

\bibitem{} Miller, J. M., Homan, J., and Miniutti, G., 2006, ApJ, 652,
  L113 (M06)

\bibitem{} Miller, J.M., Reynolds, C.S., Fabian, A.C., Cackett, E.M.,
  Miniutti, G., Raymond, J. Steeghs, D., Reis, R.C., Homan, J., 2008a, ApJ,
  679L, 113

\bibitem{} Miller, J.~M., Raymond, J., Reynolds, C.~S., Fabian, A.~C.,
  Kallman, T.~R., and Homan, J., 2008b, ApJ, 680, 1359

\bibitem{} Mitsuda, K., Inoue, H., Koyama, K., Makishima, K., Matsuoka, M.,
  Ogawara, Y., Suzuki, K., Shibazaki, N., \& Hirano, T., 1984, PASJ, 36,
  741

\bibitem{} Orosz, J. A., and Bailyn, C. D., 1997, ApJ, 477, 876

\bibitem{} Palmer, D.~M., Barthelmey, S.~D., Cummings, J.~R., Gehrels,
  N., Krimm, H.~A., Markwardt, C.~B., Sakamoto, T. and Tueller, J.,
  2005, ATEL, 546

\bibitem{} Pszota, G., and Cui, W., 2007, ApJ, 663, 1206

\bibitem{} Ramadevi, M. C., and Seetha, S., 2007, MNRAS, 378, 182

\bibitem{} Reis, R.~C., Fabian, A.~C., Ross, R.~R., Miniutti, G.,
  Miller, J.~M., and Reynolds, C., 2008, MNRAS, 689R.

\bibitem{} Reynolds, C.S., \& Begelman M. C., 1997, ApJ, 488, 109

\bibitem{} Reynolds, C.S., Fabian, A.C., 2008, ApJ, 675.1048R

\bibitem{} Reynolds, C.S., \& Nowak, M. A., 2003, PhR, 377, 389

\bibitem{} Rezzolla, L., Yoshida, S., Maccarone, T.~J., and Zanotti,
  O., 2003, MNRAS, 334L, 37

\bibitem{} Ross, R.R., \& Fabian, A.C., 1993, MNRAS, 261, 74
\bibitem{} Ross, R. R., Fabian, A. C., Young, A. J.,  1999, MNRAS, 306, 461

\bibitem{} Ross, R. R., Fabian, A. C., 2005, MNRAS, 358, 211
\bibitem{} Ross, R.R., \& Fabian, A.C., 2007, MNRAS, 381, 1697

\bibitem{} Miniutti, G. \& Fabian, A. C., 2004, MNRAS, 349, 1435


\bibitem{} Shafee, R., McClintock, J.~E., Narayan, R., Davis, S.~W., Li,
  L.-X., Remillard, R.~A., 2006, ApJ, 636, L113



\bibitem{} Soleri, P., Altamirano, D., Fender, R., Casella, P.,
  Tudose, V., Maitra, D., Wijnands, R., Belloni, T., Miller-Jones, J.,
  Klein-Wolt, M., and van der Klis, M., 2008, AIPC, 1010, 103S

\bibitem{} Struder, L., et al., 2001, A\&A, 365, L18


\bibitem{} Ueda, Y., Inoue, H., Tanaka, Y., Ebisawa, K., Nagase, F.,
  Kotani, T., and Gehrels, N., 1998, ApJ, 492, 782

\bibitem{} Wagoner, R.~V., Silbergleit, A.~S., and
  Ortega-Rodr{\'{\i}}guez, M., 2001, ApJ, 559L, 25


\bibitem{} Zhang, S.~N., Wilson, C.~A., Harmon, B~A., Fishman, G.~J.,
  Wilson, R.~B., Paciesas, W.~S., Scott, M., and Rubin, B.~C., 1994,
  IUA Circ. 6046

\bibitem{} Zhang, S.~N., Cui, W., and Chen, W., 1997, ApJ, 482, 155

\end{thebibliography}
\end{document}